\pdfoutput=1
\RequirePackage{ifpdf}
\ifpdf 
\documentclass[pdftex]{sigma}
\else
\documentclass{sigma}
\fi

\DeclareMathOperator{\sech}{sech}

\numberwithin{equation}{section}

\begin{document}

\allowdisplaybreaks

\newcommand{\arXivNumber}{1408.0588}

\renewcommand{\PaperNumber}{111}

\FirstPageHeading

\ShortArticleName{Mach-Type Soliton in the Novikov--Veselov Equation}

\ArticleName{Mach-Type Soliton in the Novikov--Veselov Equation}

\Author{Jen-Hsu CHANG}

\AuthorNameForHeading{J.H.~Chang}

\Address{Department of Computer Science and Information Engineering,\\
National Defense University, Tauyuan County 33551, Taiwan}
\Email{\href{jhchang@ndu.edu.tw}{jhchang@ndu.edu.tw}}

\ArticleDates{Received September 18, 2014, in f\/inal form December 10, 2014; Published online December 18, 2014}

\Abstract{Using the reality condition of the solutions, one constructs the Mach-type soliton of the Novikov--Veselov
equation by the minor-summation formula of the Pfaf\/f\/ian.
We study the evolution of the Mach-type soliton and f\/ind that the amplitude of the Mach stem wave is less than two times
of the one of the incident wave.
It is shown that the length of the Mach stem wave is linear with time.
One discusses the relations with $V$-shape initial value wave for dif\/ferent critical values of Miles parameter.}


\Classification{35C08; 35A22}

\section{Introduction}

Recently, the resonance theory of line solitons of KP-(II) equation (shallow water wave equation)
\begin{gather*}
\partial_x (-4 u_t+u_{xxx}+6uu_x)+3u_{yy}=0
\end{gather*}
has attracted much attractions using the totally non-negative Grassmannian~\cite{ad, bc,ko, ch, ko1, ko4, ko3}, that
is, those points of the real Grassmannian whose Plucker coordinates are all non-negative.
For the KP-(II) equation case, the~$\tau$-function is described by the Wronskian form with respect to~$x$.
The Mach ref\/lection problem describes the resonant interaction of solitary waves appearing in the ref\/lection of an
obliquely incident waves onto a~vertical wall.
John Miles discussed an oblique interaction of solitary waves and found a~resonant interaction to describe the Mach
ref\/lection phenomena~\cite{mi}.
In this end, he predicts an extraordinary fourfold application of the wave at the wall.
The Miles theory in terms of the KP equation and the Mach-type solution in KP observed experimentally are investigated
in~\cite{ko1, ko5,ko6, koo, ye} (and references therein).
The point is that irregular ref\/lection can be described by the (3142)-type soliton and the stem in the middle part
should be a~Mach stem wave.
Inspired by their works, one can consider the Novikov--Veselov equation similarly.

One considers the Novikov--Veselov (NV)
equation~\cite{ba,gm,kr,nv,vn} with real solution~$U$:
\begin{gather}
U_t=\operatorname{Re} \big[\partial_z^3 U+3\partial_z(QU)-3 \epsilon {\partial}_z Q\big],
\label{NV}
\\
{\bar{\partial}}_z Q=\partial_z U,
\qquad
t \in R,
\nonumber
\end{gather}
where~$\epsilon$ is a~real constant.
The NV equation~\eqref{NV} is one of the natural generation of the famous KdV equation and can have the Manakov's triad
representation~\cite{ma}
\begin{gather*}
L_t=[A, L]+BL,
\end{gather*}
where~$L$ is the two-dimension Schr\"odinger operator
\begin{gather*}
L=\partial_z \bar{\partial_z} +U- \epsilon
\end{gather*}
and
\begin{gather*}
A=\partial_z^3+Q\partial_z+{\bar{\partial}}_z^3+\bar{Q}{\bar{\partial}}_z,
\qquad
B=Q_z+\bar{Q}_{\bar{z}}.
\end{gather*}
It is equivalent to the linear representation
\begin{gather}
L \phi=0,
\qquad
\partial_t \phi=A \phi.
\label{rep}
\end{gather}
We remark that when $\epsilon \to \pm \infty$, the Veselov--Novikov equation reduces to the KP-I ($\epsilon \to - \infty
$) and KP-(II) ($\epsilon \to \infty$) equation respectively~\cite{gri}.
To make a~comparison with KP-(II) equation, we only consider $\epsilon>0$.

Let $\phi_1$, $\phi_2$ be any two independent solutions of~\eqref{rep}.
Then one can construct the extended Moutard transformation using the skew product~\cite{an, hh, ms}
\begin{gather}
W(\phi_1, \phi_2)=\int (\phi_1\partial \phi_2-\phi_2\partial \phi_1)dz-\big(\phi_1\bar{\partial}
\phi_2-\phi_2\bar{\partial} \phi_1\big)d \bar{z}+\big[\phi_1\partial^3 \phi_2-\phi_2\partial^3 \phi_1
\nonumber
\\
\phantom{W(\phi_1, \phi_2)=}{}
+\phi_2\bar{\partial}^3 \phi_1-\phi_1\bar{\partial}^3 \phi_2+2\big(\partial^2\phi_1 \partial
\phi_2-\partial\phi_1\partial^2\phi_2\big)-2\big(\bar{\partial}^2\phi_1\bar{\partial}\phi_2-\bar{\partial}\phi_1
\bar{\partial}^2\phi_2\big)
\nonumber
\\
\phantom{W(\phi_1, \phi_2)=}{}
+3 Q (\phi_1\partial \phi_2-\phi_2\partial \phi_1)-3\bar{Q}\big(\phi_1\bar{\partial} \phi_2-\phi_2\bar{\partial}
\phi_1\big)\big]dt,
\label{ext}
\end{gather}
such that
\begin{gather*}
\hat{U}(t,z,\bar{z})=U(t,z,\bar{z})+2\partial \bar{\partial} \ln W(\phi_1, \phi_2),
\qquad
\hat{Q} (t,z,\bar{z})=Q(t,z,\bar{z})+2 \partial \partial \ln W (\phi_1, \phi_2),
\end{gather*}
is also a~solution of the NV equation~\eqref{NV}.

For f\/ixed potential $U_0(z, \bar{z}, t)$ and $Q_0(z, \bar{z}, t)$ of the NV equation~\eqref{NV}, we can take any $2N$
wave functions $\phi_1, \phi_2, \phi_3, \dots, \phi_{2N}$ (or their linear combinations) of~\eqref{rep}.
Then the $2N$-step successive extended Moutard transformation can be expressed as the Pfaf\/f\/ian form~\cite{an, ni} (also
see~\cite{he, oh})
\begin{gather*}
U=U_0+2 \partial \bar{\partial} [\ln \operatorname{Pf}(\phi_1, \phi_2, \phi_3, \dots, \phi_{2N})],
\qquad
Q=Q_0+2 \partial \partial [\ln \operatorname{Pf}(\phi_1, \phi_2, \phi_3, \dots, \phi_{2N})],
\end{gather*}
where $Pf(\phi_1, \phi_2, \phi_3, \dots,\phi_{2N})$ is the Pfaf\/f\/ian def\/ined by
\begin{gather*}
\operatorname{Pf}(\phi_1, \phi_2, \phi_3, \dots, \phi_{2N})=\sum\limits_{\sigma}\epsilon(\sigma)W_{\sigma_1 \sigma_2}W_{\sigma_3
\sigma_4} \cdots W_{\sigma_{2N-1} \sigma_{2N}},
\end{gather*}
and $W_{\sigma_{i} \sigma_j}=W(\phi_{\sigma(i)},\phi_{\sigma(j)})$ is the extended Moutard
transformation~\eqref{ext},~$\sigma$ being some permutations.

To construct the~$N$-solitons solutions, we take $V=U=0$ in~\eqref{NV} and then~\eqref{rep} becomes
\begin{gather}
\partial \bar{\partial} \phi=\epsilon \phi,
\qquad \phi_t=\phi_{zzz}+\phi_{\bar{z}\bar{z}\bar{z}},
\label{pri}
\end{gather}
where~$\epsilon$ is non-zero real constant.
The general solution of~\eqref{pri} can be expressed as
\begin{gather}
\phi(z, \bar{z}, t)=\int_{\Gamma} e^{(i\lambda) z+(i\lambda)^3
t+\frac{\epsilon}{i\lambda}\bar{z}+\frac{\epsilon^3}{(i\lambda)^3}t} \nu(\lambda) d \lambda,
\label{con}
\end{gather}
where $\nu(\lambda)$ is an arbitrary distribution and~$\Gamma$ is an arbitrary path of integration such that the r.h.s.\
of~\eqref{con} is well def\/ined.
One takes $\nu_m(\lambda)=\delta (\lambda-p_m)$, where $p_m$ is a~complex number.
Def\/ine
\begin{gather*}
\phi_m=\frac{\phi (p_m)}{\sqrt{3}}=\frac{1}{\sqrt{3}}e^{F(p_m)},
\end{gather*}
where
\begin{gather*}
F(\lambda)=(i\lambda)z+(i\lambda)^3t+\frac{\epsilon}{i\lambda}\bar{z}+\frac{\epsilon^3}{(i\lambda)^3}t.
\end{gather*}
Plugging $(\phi_m,\phi_n)$ into the extended Moutard transformation~\eqref{ext}, we obtain
\begin{gather}
W(\phi_m,\phi_n)=i \frac{p_n-p_m}{p_n+p_m} e^{F(p_m)+F(p_n)}.
\label{bac}
\end{gather}

To study resonance, we introduce the real Grassmannian (or the $2N \times M$ matrix) to construct $N$~solitons.
To this end, one considers linear combination of~$\phi_n$.
Let
\begin{gather*}
\vec{\Psi}=(\phi_1, \phi_2, \phi_3, \dots, \phi_M)^T
\end{gather*}
and~$H$ be an $2N \times M$ $(2N \leq M)$ of real constant matrix (or Grassmannian).
Suppose that
\begin{gather*}
H \vec{\Psi}=\vec{\Psi}^\ast=(\Psi_1^\ast, \Psi_2^\ast, \Psi_3^\ast, \dots, \Psi_{2N}^\ast)^T,
\end{gather*}
that is,
\begin{gather*}
\Psi_n^\ast=h_{n1} \phi_1+h_{n2} \phi_2+h_{n3} \phi_3+\dots +h_{nM} \phi_M,
\qquad
1\leq n \leq 2N.
\end{gather*}
Then one has by the minor-summation formula~\cite{is, km}
\begin{gather}
\tau_N=\operatorname{Pf}(\Psi_1^\ast, \Psi_2^\ast, \Psi_3^\ast, \dots, \Psi_{2N}^\ast)=\operatorname{Pf}\big(H W_M H^T\big)
=\sum\limits_{I\subset [M],\;
\sharp I=2N} \operatorname{Pf}\big(H_I^I\big)\det (H_I),
\label{sum}
\end{gather}
where the $M \times M$ matrix $W_M$ is def\/ined by the element~\eqref{bac} and $H_I^I$ denote the $2N \times M$
submatrix of~$H$ obtained by picking up the rows and columns indexed by the same index set~$I$.
By this formula, the resonance of real solitons of the Novikov--Veselov equation can be investigated just like the
resonance theory of KP-(II) equation~\cite{ko1, ko2, ko3}.
Finally, the $N$-solitons solutions are def\/ined by~\cite{jh1, bd}
\begin{gather*}
U(z, \bar{z}, t)=2 \partial \bar{\partial} \ln \tau_N (z, \bar{z}, t),
\qquad
V(z, \bar{z}, t)=2 \partial \partial \ln \tau_N (z, \bar{z}, t).
\end{gather*}
To obtain the real potential~$U$, the following reality conditions~\cite{bd} for resonance have to be considered
\begin{gather*}
|p_k|^2=|q_k|^2=\epsilon>0,
\qquad
k=1,2,3, \dots, m,
\end{gather*}
given~$m$ pairs of complex numbers $(p_1, q_1), (p_2, q_2), \dots,(p_m, q_m)$.

Letting $p_m=\sqrt{\epsilon} e^{i \alpha_m}$ and {\it removing~$i$ factor from~\eqref{bac} afterwards}, one has
\begin{gather}
W(\phi_m,\phi_n)=-\tan\frac{\alpha_n-\alpha_m}{2} e^{\phi_{mn}},
\label{pp}
\end{gather}
where
\begin{gather}
\phi_{mn}=F(p_m)+F(p_n)=-2\sqrt{\epsilon}[x(\sin\alpha_m+\sin\alpha_n)+y(\cos\alpha_m+\cos \alpha_n)]
\nonumber
\\
\phantom{\phi_{mn}=}{}
+2t\epsilon \sqrt{\epsilon}(\sin3\alpha_m+\sin3\alpha_n).
\label{phi}
\end{gather}
Therefore, given a~$2N \times M$ matrix~$H$, the associated $\tau_H$-function can be written as
by~\eqref{sum}~\cite{jh2}
\begin{gather*}
\tau_H=\sum\limits_{I \subset [M],\;
\sharp I=2N} \Gamma_I \Lambda_I(x,y,t),
\end{gather*}
where
\begin{gather*}
\Lambda_I(x,y,t)=\operatorname{Pf} (W_{2N})
=(-1)^N \left (\prod\limits_{i=2,\, i> j}^{2N} \tan \frac{\alpha_i-\alpha_j}{2}\right)e^{\sum\limits_{m=1}^{2N} F(p_m)},
\end{gather*}
$\Gamma_I$ being the $2N \times 2N$ minor for the columns with the index set $I=\{i_1, i_2, i_3, \dots, i_{2N}\}$.
Also, to keep $\tau_H$ totally positive (or totally negative), we assume that the matrix~$H$ belongs to the totally
non-negative Grassmannian~\cite{ko4, ko3} and the angle $\alpha_n$ satisf\/ies the following condition:
\begin{gather*}
-\frac{\pi}{2} \leq \alpha_1< \alpha_2 <\alpha_3<\dots<\alpha_{M-1}<\alpha_{M} \leq \frac{\pi}{2}.
\end{gather*}
For one-soliton solution, we have, $-\frac{\pi}{2} \leq \alpha_i< \alpha_j <\alpha_k \leq \frac{\pi}{2}$
\begin{gather*}
\tau_1=\tan \frac{\alpha_i-\alpha_j}{2} e^{\phi_{ij}}+a\tan \frac{\alpha_i-\alpha_k}{2} e^{\phi_{ik}}
\\
\phantom{\tau_1}{}
=ae^{\phi_{ik}} \tan \frac{\alpha_i-\alpha_k}{2}\left[1+\frac{1}{a} \frac{\tan \frac{\alpha_i-\alpha_j}{2}}{\tan
\frac{\alpha_i-\alpha_k}{2}} e^{F(p_j)-F(p_k)}\right]
\\
\phantom{\tau_1}{}
=ae^{\phi_{ik}} \tan \frac{\alpha_i-\alpha_k}{2}\big[1+e^{F(p_j)-F(p_k)+\theta_{jk}}\big],
\end{gather*}
where~$a$ is a~constant and the phase shift
\begin{gather*}
\theta_{jk}=\ln \frac{1}{a} \frac{\tan \frac{\alpha_i-\alpha_j}{2}}{\tan \frac{\alpha_i-\alpha_k}{2}}=\ln \frac{\tan
\frac{\alpha_i-\alpha_j}{2}}{\tan \frac{\alpha_i-\alpha_k}{2}}-\ln a.
\end{gather*}
Hence the real one-soliton solution is~\cite{jh2}
\begin{gather}
U=2 \partial_{z}\partial_{\bar{z}} \ln ae^{\phi_{ik}} \tan \frac{\alpha_i-\alpha_k}{2}\big[1+e^{F(p_j)-F(p_k)+\theta_{jk}}\big]
=2 \partial_{z}\partial_{\bar{z}}\big[1+e^{F(p_j)-F(p_k)+\theta_{jk}}\big]
\nonumber
\\
\phantom{U}
=\frac{1}{2}\big|p_k-p_j\big|^2 \sech^2 \left[\frac{F(p_j)-F(p_k)+\theta_{jk}}{2}\right]
\nonumber
\\
\phantom{U}
=2 \epsilon \sin^2\left(\frac{\alpha_k-\alpha_j}{2}\right)\sech^2\left[\frac{F(p_j)-F(p_k)+\theta_{jk}}{2}\right]
\nonumber
\\
\phantom{U}
=A_{[j,k]} \sech^2 \frac{1}{2} \big(\vec{\bf{K}}_{[j,k]} \cdot \vec{x}-{\bf{\Omega}}_{[j,k]}t+\theta_{jk} \big).
\label{on}
\end{gather}
From~\eqref{phi} the amplitude $A_{[j,k]}$, the wave vector $\vec{\bf{K}}_{[j,k]}$ and the frequency
${\bf\Omega}_{[j,k]}$ are def\/ined~by
\begin{gather}
A_{[j,k]}=2 \epsilon \sin^2\left(\frac{\alpha_k-\alpha_j}{2}\right),
\nonumber
\\
\vec{\bf{K}}_{[j,k]}=2 \sqrt{\epsilon} (-\sin \alpha_j+\sin \alpha_k, -\cos \alpha_j+\cos \alpha_k),
\nonumber
\\
{\bf\Omega}_{[j,k]}=2 \epsilon \sqrt{\epsilon} [-\sin 3 \alpha_j+\sin 3\alpha_k],\label{am}
\end{gather}
The direction of the wave vector $\vec{\bf{K}}_{[j,k]}=\big(K_{[j,k]}^x, K_{[j,k]}^y\big)$ is measured in the clockwise sense
from the $y$-axis and it is given~by
\begin{gather*}
\frac{K_{[j,k]}^y}{K_{[j,k]}^x}=\frac{-\cos \alpha_j+\cos \alpha_k}{-\sin \alpha_j+\sin \alpha_k}=-\tan\frac{\alpha_j+\alpha_k}{2},
\end{gather*}
that is, $\frac{\alpha_j+\alpha_k}{2}$ gives the angle between the line soliton and the $y$-axis in the clockwise sense.
In addition, the soliton velocity $\bf{V_{[j,k]}}$ is~\cite{jh2}
\begin{gather}
{\bf{V_{[j,k]}}}=\frac{\epsilon}{4} \frac{\sin 3 \alpha_k-\sin 3 \alpha_j}{\sin^2\frac{\alpha_j-\alpha_k}{2}}
\left(\sin\alpha_k-\sin \alpha_j, \cos \alpha_k-\cos \alpha_j\right).
\label{v}
\end{gather}

The paper is organized as follows.
In Section~\ref{section2}, one investigates Mach-type or (3142)-type soliton for the Novikov--Veselov equation.
One shows the evolution of the Mach-type soliton and obtains the relation of the amplitude of the Mach stem wave
([1,4]-soliton) with the one of the incident wave ([1,3]-soliton).
Furthermore, the length of the Mach stem wave is linear with time.
In Section~\ref{section3}, we discuss the relations with $V$-shape initial value wave for dif\/ferent critical value of
Miles parameter~$\kappa$.
It is shown that the amplitude of the Mach stem wave is less than two times of the one of the incident wave.
In Section~\ref{section4}, we conclude the paper with several remarks.

\section{Mach type soliton}
\label{section2}

In this section, we investigate the Mach-type or (3142)-type soliton.
The corresponding totally non-negative Grassmannian is the the matrix~\cite{ko1}
\begin{gather*}
H_M=\left[
\begin{matrix}
1 & a & 0 & -c
\\
0 & 0 & 1 & b
\end{matrix}
\right],
\end{gather*}
where $a$, $b$, $c$ are positive numbers.
When $c=0$, one has the~$O$-type soliton for the Novikov--Veselov equation.
For $V$-shape initial value wave, one can introduce parameter~$\kappa$ to determine the evolution into Mach-type
or~$O$-type soliton (see next section).
We remark that the~$Y$-shape,~$O$-type, and~$P$-type solitons for the Novikov--Veselov equation are investigated
in~\cite{jh2}.
Now,
\begin{gather*}
H_M \frac{1}{\sqrt{3}}[\phi(p_1), \phi(p_2), \phi(p_3), \phi(p_4)]^T=\frac{1}{\sqrt{3}}\left[
\begin{matrix} \phi(p_1)+a \phi(p_2)
\\
\phi(p_3)+b\phi(p_4)
\end{matrix}
\right]=\left[
\begin{matrix} \Psi_1^\ast
\\
\Psi_2^\ast
\end{matrix}
\right].
\end{gather*}
A~direct calculation yields by~\eqref{sum},~\eqref{pp} and~\eqref{phi}
\begin{gather}
\tau_M=W(\Psi_1^\ast,\Psi_2^\ast)=W(\phi_1, \phi_3)+b W(\phi_1, \phi_4)+a W(\phi_2, \phi_3)+ab W(\phi_2, \phi_4)+c W(\phi_3,\phi_4)
\nonumber
\\
\phantom{\tau_M}
=\tan{\frac{\alpha_1-\alpha_3}{2}} e^{-2\sqrt{\epsilon}[x(\sin\alpha_1+\sin\alpha_3)+y(\cos\alpha_1+\cos \alpha_3)]
+2t\epsilon \sqrt{\epsilon}(\sin3\alpha_1+\sin3\alpha_3)}
\nonumber
\\
\phantom{\tau_M=}
+b \tan{\frac{\alpha_1-\alpha_4}{2}} e^{-2\sqrt{\epsilon}[x(\sin\alpha_1+\sin\alpha_4)+y(\cos\alpha_1+\cos
\alpha_4)]+2t\epsilon \sqrt{\epsilon}(\sin3\alpha_1+\sin3\alpha_4)}
\nonumber
\\
\phantom{\tau_M=}
+a\tan{\frac{\alpha_2-\alpha_3}{2}} e^{-2\sqrt{\epsilon}[x(\sin\alpha_2+\sin\alpha_3)+y(\cos\alpha_2+\cos
\alpha_3)]+2t\epsilon \sqrt{\epsilon}(\sin3\alpha_2+\sin3\alpha_3)}
\nonumber
\\
\phantom{\tau_M=}
+ab \tan{\frac{\alpha_2-\alpha_4}{2}} e^{-2\sqrt{\epsilon}[x(\sin\alpha_2+\sin\alpha_4)+y(\cos\alpha_2+\cos
\alpha_4)]+2t\epsilon \sqrt{\epsilon}(\sin3\alpha_2+\sin3\alpha_4)}
\nonumber
\\
\phantom{\tau_M=}
+c \tan{\frac{\alpha_3-\alpha_4}{2}} e^{-2\sqrt{\epsilon}[x(\sin\alpha_3+\sin\alpha_4)+y(\cos\alpha_3+\cos
\alpha_4)]+2t\epsilon \sqrt{\epsilon}(\sin3\alpha_3+\sin3\alpha_4)},
\label{oo}
\end{gather}
where
\begin{gather*}
-\frac{\pi}{2} \leq \alpha_1< \alpha_2 <\alpha_3<\alpha_{4} \leq \frac{\pi}{2}.
\end{gather*}
To investigate the asymptotic behavior for $\vert y \vert \to \infty$, we use the notation~\cite{ko1}, considering the
line $x=-cy$,
\begin{gather*}
\eta_m (c)=-c \sin \alpha_m+\cos \alpha_m.
\end{gather*}
When $\eta_m (c)=\eta_n (c)$, one gets
\begin{gather*}
c=\frac{\cos \alpha_m-\cos \alpha_n}{\sin \alpha_m -\sin \alpha_n}=-\tan \frac{\alpha_m+\alpha_n}{2}.
\end{gather*}
Since
\begin{gather*}
[\eta_m (c)- \eta_i (c)] \Big\vert_{c=-\tan \frac{\alpha_i+\alpha_j}{2}}
=\cos \alpha_m-\cos \alpha_i+\tan \frac{\alpha_i+\alpha_j}{2} (\sin \alpha_m -\sin \alpha_i)
\\
\hphantom{[\eta_m (c)- \eta_i (c)] \Big\vert_{c=-\tan \frac{\alpha_i+\alpha_j}{2}}}{}
=(\sin \alpha_m -\sin \alpha_i) \left[\tan \frac{\alpha_i+\alpha_j}{2}-\tan \frac{\alpha_i+\alpha_m}{2}\right],
\end{gather*}
we have the following order relations among the other $\eta_m(c)'s $
along $c=-\tan \frac{\alpha_i+\alpha_j}{2}$
\begin{gather*}
\begin{cases}
\eta_i=\eta_j<\eta_m
&
\text{if}
\quad
i< m<j,
\\
\eta_i=\eta_j>\eta_m
&
\text{if}
\quad
m<i
\quad
\text{or}
\quad
m >j.
\end{cases}
\end{gather*}
Then by a~similar argument in~\cite{ko1}, one knows that by~\eqref{on}:

(a) For $y \gg 0$, there are two unbounded line solitons, whose types from left to right are
\begin{gather*}
[1,3],
\qquad
[3,4].
\end{gather*}

(b) For $y \ll 0$, there are two unbounded line solitons, whose types from left to right are
\begin{gather*}
[4,2],
\qquad
[2,1].
\end{gather*}
It can be verif\/ied by the Maple software.

\begin{figure}[t] \centering
\includegraphics[width=0.50\textwidth]{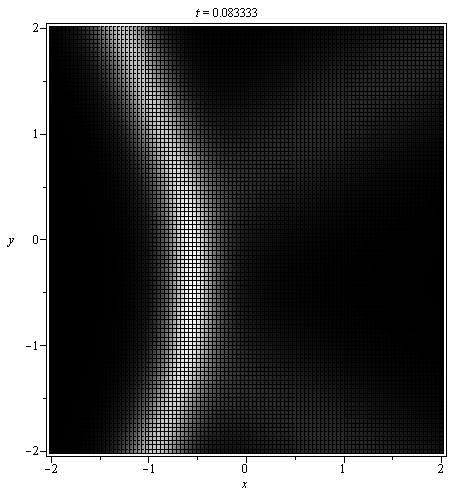}
\caption{The middle portion, having maximum amplitude, is the
[1,4]-soliton (stem wave). The $y$-axis is slightly enlarged to make the middle portion longer.}\label{Fig1}
\end{figure}

\begin{figure}[t!] \centering
\includegraphics[width=0.5\textwidth]{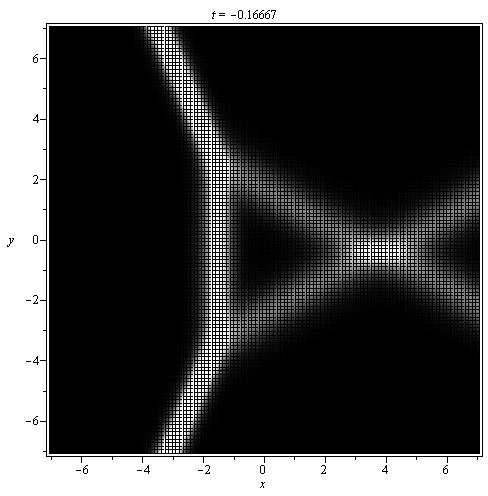}
\caption{$\alpha_1=-\frac{23}{50}\pi$,
$\alpha_2=-\frac{1}{5}\pi$, $\alpha_3=\frac{1}{5}\pi$, $\alpha_4=\frac{23}{50}\pi$, $a=b=c=1$, $\epsilon=5$.}\label{Fig2}
\end{figure}

Now, we can discuss the relations between the parameters $a$, $b$, $c$ and phase shifts of these line solitons.
Let us f\/irst consider the line solitons in $x \gg 0$.
There are two solitons which are [3,4]-soliton and [2,1]-soliton.
The [3,4]-soliton is obtained by the balance between the exponential terms $ W(\phi_1, \phi_3)$ and $ b W(\phi_1,
\phi_4)$, and the [2,1]-soliton is obtained by the balance between the exponential terms $ W(\phi_1, \phi_3)$ and $aW(\phi_2, \phi_3)$.
Therefore, the phase shifts of [3,4]-soliton and [2,1]-soliton for $x\gg 0$ are given~by
\begin{gather*}
\theta_{[3,4]}^+=\ln \frac{\tan \frac{\alpha_3-\alpha_1}{2}}{\tan \frac{\alpha_4-\alpha_1}{2}} -\ln b,
\qquad
\theta_{[2,1]}^+=\ln \frac{\tan \frac{\alpha_3-\alpha_1}{2}}{\tan \frac{\alpha_3-\alpha_2}{2}} -\ln a.
\end{gather*}
For the line solitons in $x\ll 0$, there are two solitons, which are [1,3]-soliton and [4,2]-soliton.
The [1,3]-soliton is obtained by the balance between the exponential terms~$c W(\phi_3, \phi_4)$ and $ b W(\phi_1,
\phi_4)$, and the [4,2]-soliton is obtained by the balance between the exponential terms $ c W(\phi_3, \phi_4)$ and $aW(\phi_2, \phi_3)$.
Therefore, the phase shifts of [1,3]-soliton and [4,2]-soliton for $x\ll 0$ are given~by
\begin{gather*}
\theta_{[1,3]}^-=\ln \frac{\tan \frac{\alpha_4-\alpha_1}{2}}{\tan \frac{\alpha_4-\alpha_3}{2}} +\ln \frac{b}{c},
\qquad
\theta_{[4,2]}^-=\ln \frac{\tan \frac{\alpha_3-\alpha_2}{2}}{\tan \frac{\alpha_4-\alpha_3}{2}} +\ln \frac{a}{c}.
\end{gather*}
So one can see that
\begin{gather*}
\theta_{[1,3]}^-+\theta_{[3,4]}^+=\theta_{[4,2]}^-+\theta_{[2,1]}^+=\text{total phase shift}
=\ln \frac{\tan\frac{\alpha_3-\alpha_1}{2}}{\tan \frac{\alpha_4-\alpha_3}{2}} -\ln c.
\end{gather*}
We def\/ine the parameter~$s$ (representing the total phase shift)~by
$s=e^{-\theta_{[4,2]}^- - \theta_{[2,1]}^+}$,
which leads to
\begin{gather*}
a=\frac{\tan \frac{\alpha_3-\alpha_1}{2}}{\tan \frac{\alpha_3-\alpha_2}{2}} s e^{\theta_{[4,2]}^-},
\qquad
b=\frac{\tan \frac{\alpha_3-\alpha_1}{2}}{\tan \frac{\alpha_4-\alpha_1}{2}} s e^{\theta_{[1,3]}^-},
\qquad
c=\frac{\tan \frac{\alpha_3-\alpha_1}{2}}{\tan \frac{\alpha_4-\alpha_3}{2}} s.
\end{gather*}
Hence we know that the three parameters $a$, $b$, $c$ can be used to determine the locations
of three asymptotic line
solitons, that is, two in $x\ll 0$ and one in $x >0$.
The~$s$-parameter represents the relative locations of the intersection point of the [1,3]-soliton and [3,4]-soliton with the $x$-axis.
Especially, when $ s=1$, $\theta_{[4,2]}^-=0$, $\theta_{[1,3]}^-=0$, all of the four solitons will intersect at $(0,0)$ when $t=0$.
One remarks that the bounded line soliton [1,4] (Mach stem wave), obtained by the balance between the exponential terms
$W(\phi_1, \phi_3)$ and $c W(\phi_3, \phi_4)$, has the maximal amplitude among all the solitons by~\eqref{am} (Fig.~\ref{Fig1})
and the velocity is obtained by~\eqref{v}.
Furthermore, when $t<0$, there is a~bounded line [2,3]-soliton (Fig.~\ref{Fig2},
the left side of the triangle), obtained~by
the balance between the exponential terms $ ab W(\phi_2, \phi_4)$ and $c W(\phi_3, \phi_4)$.

\begin{figure}[t]
\centering
\includegraphics[width=0.5\textwidth]{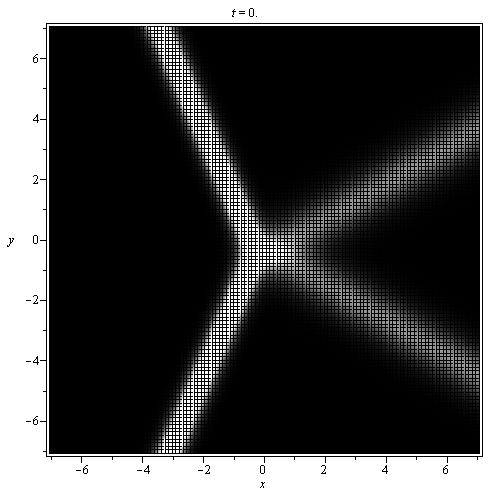} \caption{Initial wave.}\label{Fig3}
\end{figure}

Now, we consider the case $\alpha_3=-\alpha_2 \geq 0$, $\alpha_4=-\alpha_1 \geq 0$, and the amplitude
\begin{gather}
A=A_{[1,3]}=A_{[4,2]} \leq 2 \epsilon
\label{A1}
\end{gather}
is f\/ixed.
Then one can see that $[1,3]$-soliton and $[4,2]$-soliton is symmetric to the $x$-axis and similarly for $[3,4]$-soliton
and $[2,1]$-soliton.
By~\eqref{am}, one knows
\begin{gather*}
\frac{\alpha_3+\alpha_1}{2} \leq \frac{\alpha_3 - \alpha_1}{2}=\frac{\alpha_3+\alpha_4}{2}=\arcsin
\sqrt{\frac{A}{2\epsilon}}.
\end{gather*}
Therefore the angle between the [1,3]-soliton and the $y$-axis (counter-clockwise) is less than the the angle between the
[3,4]-soliton and the $y$-axis (clockwise).
We see that given~$A$ and $2 \epsilon $ there is a~critical
angle $\varphi_C=\arcsin \sqrt{\frac{A}{2\epsilon}}$ for the
angle between the [1,3]-soliton and the $y$-axis (counter-clockwise).
Then one can introduce the following Miles-parameter~\cite{ko,jh2, mi} to describe the interaction for the Mach-type
solution, noticing $\frac{\alpha_3+\alpha_1}{2} \leq 0$,
\begin{gather}
\kappa=\frac{\vert \tan \frac{\alpha_3+\alpha_1}{2} \vert}{\tan \frac{\alpha_3+\alpha_4}{2}}=\frac{\vert \tan
\frac{\alpha_3+\alpha_1}{2} \vert}{\tan \varphi_C}=\frac{\vert \tan \frac{\alpha_3+\alpha_1}{2}
\vert}{\sqrt{\frac{A}{2\epsilon-A}}} \leq 1.
\label{ka}
\end{gather}
From~\eqref{am}, we have thus using $\kappa$
\begin{gather}
A=A_{[1,3]}=A_{[4,2]}=\frac{2\epsilon (\tan \varphi_C)^2}{1+(\tan \varphi_C)^2},
\qquad
A_{[3,4]}=A_{[2,1]}=\frac{2\epsilon (\tan \varphi_C)^2}{\frac{1}{\kappa^2}+(\tan \varphi_C)^2} \leq A,
\nonumber
\\
A_{[1,4]}=2\epsilon \sin^2 \frac{\alpha_4-\alpha_1}{2}
=2\epsilon\left[\sin\left(\frac{\alpha_4-\alpha_3}{2}+\frac{\alpha_3-\alpha_1}{2}\right)\right]^2
\nonumber
\\
\phantom{A_{[1,4]}}
=\frac{2\epsilon (\tan \varphi_C)^2 (\kappa+1)^2}{[1+(\tan \varphi_C)^2] [1+\kappa^2 (\tan \varphi_C)^2]}
=A\frac{(\kappa+1)^2}{[1+\kappa^2 (\tan \varphi_C)^2]}<4A.
\label{a}
\end{gather}

\begin{remark*}
 To make a~comparison with KP-(II), we see that
\begin{gather*}
2 \epsilon -A=2 \epsilon\left(1- \sin^2 \frac{\alpha_3-\alpha_1}{2}\right)=2 \epsilon \cos^2 \frac{\alpha_3-\alpha_1}{2}.
\end{gather*}
When $\epsilon \to \infty$, $\alpha_1 \to -\frac{\pi}{2}$ and $\alpha_3 \to \frac{\pi}{2}$ such that
\begin{gather*}
\epsilon \cos^2 \frac{\alpha_3-\alpha_1}{2}=\frac{1}{4}.
\end{gather*}
Then
\begin{gather*}
\kappa \to \frac{\vert \tan \frac{\alpha_3+\alpha_1}{2} \vert}{\sqrt{2A}},
\end{gather*}
which is the Miles parameter of KP-(II)
to describe the interactions of water wave solitons~\cite{ ko1, ko5, ko6}.
\end{remark*}

\begin{figure}[t] \centering
\includegraphics[width=0.50\textwidth]{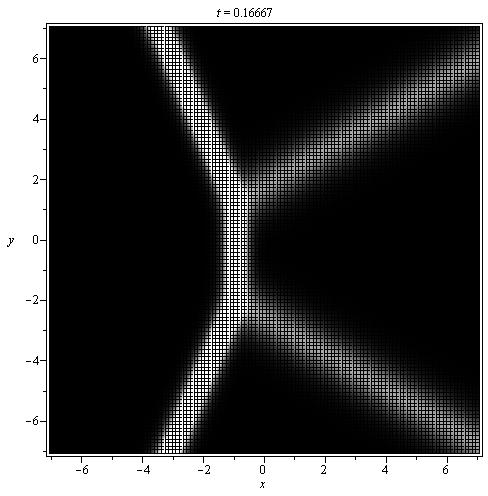}
\caption{The stem wave moves to the left.}\label{Fig4}
\end{figure}

\begin{figure}[th!] \centering
\includegraphics[width=0.50\textwidth]{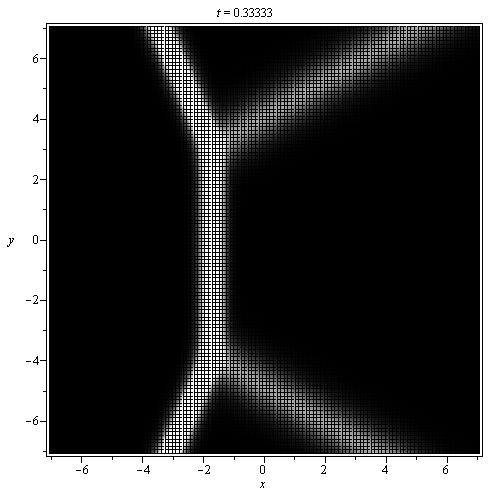}
\caption{The length of the stem wave is increasing.}\label{Fig5}
\end{figure}

Since the [1,4]-soliton (Mach stem wave) is increasing its length with time but its end points will lie in a~line (see
Figs.~\ref{Fig4} and~\ref{Fig5}), we can obtain them as follows.

We choose $ s=1$, $\theta_{[4,2]}^-=0$, $\theta_{[1,3]}^-=0 $ such that [1,3]-soliton and [1,4]-soliton will intersect at
$(0,0)$ when $t=0$ (see Fig.~\ref{Fig3}).
From~\eqref{on}), the ridges of [1,3]-soliton and [1,4]-soliton are given by $F(p_1)-F(p_3)=0$,
$F(p_1)-F(p_4)=0$, which lead to
\begin{gather*}
 x(-\sin\alpha_1+\sin \alpha_3)+y(-\cos \alpha_1+\cos \alpha_3)+t \epsilon (\sin 3 \alpha_1- \sin 3 \alpha_3)=0,
\\
 x(-\sin\alpha_1+\sin \alpha_4)+y(-\cos \alpha_1+\cos \alpha_4)+t \epsilon (\sin 3 \alpha_1- \sin 3 \alpha_4)=0.
\end{gather*}
Noticing that $\alpha_3=-\alpha_2 \geq 0$, $\alpha_4=-\alpha_1 \geq 0$, one gets
\begin{gather}
x=\frac{t \epsilon \sin 3 \alpha_4}{\sin \alpha_4}=t\epsilon (4 \cos^2 \alpha_4 -1),
\label{x}
\\
y=\frac{x(\sin\alpha_1- \sin \alpha_3)+t \epsilon (-\sin 3 \alpha_1+\sin 3 \alpha_3)}{-\cos \alpha_1+\cos
\alpha_3}
\nonumber
\\
\phantom{y}
=t\epsilon \frac{(4 \cos^2 \alpha_4 -1) (\sin\alpha_1- \sin \alpha_3)+(-\sin 3 \alpha_1+\sin 3 \alpha_3)}{-\cos
\alpha_1+\cos \alpha_3}
\label{y}
\\
\phantom{y}
=4t \epsilon \frac{\sin\alpha_3 (\sin\alpha_3+\sin \alpha_4) (\sin\alpha_4- \sin \alpha_3)}{\cos \alpha_3- \cos
\alpha_4}
=4t \epsilon \sin\alpha_3 (\sin\alpha_3+\sin \alpha_4) \cot \frac{\alpha_3+\alpha_4}{2}.
\nonumber
\end{gather}
Using~\eqref{ka}, one has
\begin{gather}
\sin\alpha_3=\sin\left(\frac{\alpha_3+\alpha_4}{2}-\frac{\alpha_4-\alpha_3}{2}\right)
=\frac{(1-\kappa)\tan\varphi_C}{\sqrt{[1+(\tan \varphi_C)^2] [1+\kappa^2 (\tan \varphi_C)^2]}},
\nonumber
\\
\cos \alpha_4=\cos\left(\frac{\alpha_3+\alpha_4}{2}+\frac{\alpha_4-\alpha_3}{2}\right)
=\frac{1-\kappa (\tan\varphi_C)^2}{\sqrt{[1+(\tan \varphi_C)^2] [1+\kappa^2 (\tan \varphi_C)^2]}},
\nonumber
\\
\sin\alpha_3 +\sin \alpha_4=2 \sin \frac{\alpha_3+\alpha_4}{2} \cos \frac{\alpha_3-\alpha_4}{2}
=\frac{2\tan\varphi_C}{\sqrt{[1+(\tan \varphi_C)^2] [1+\kappa^2 (\tan \varphi_C)^2]}},
\nonumber
\\
4\cos^2\alpha_4-1=\frac{4 [1-\kappa (\tan \varphi_C)^2]^2}{[1+(\tan \varphi_C)^2][1+\kappa^2(\tan\varphi_C)^2]} -1
\nonumber
\\
\phantom{4\cos^2\alpha_4-1}
=\frac{3+(\tan \varphi_C)^2 [3 \kappa^2 (\tan \varphi_C)^2 -\kappa^2- 8 \kappa-1]}{[1+(\tan \varphi_C)^2][1+\kappa^2 (\tan \varphi_C)^2]}.
\label{co}
\end{gather}
A~simple calculation yields using~\eqref{a}
\begin{gather*}
y=8t \epsilon \tan \varphi_C \frac{1-\kappa}{[1+(\tan \varphi_C)^2] [1+\kappa^2 (\tan \varphi_C)^2]}=4t
A_{[1,4]} \frac{1-\kappa}{(1+\kappa^2) (\tan \varphi_C)},
\\
\tan \chi=\frac{y}{x}=\frac{8(1-\kappa)\tan \varphi_C}{3+(\tan \varphi_C)^2 [3 \kappa^2 (\tan \varphi_C)^2
-\kappa^2- 8 \kappa-1]}.
\end{gather*}
Hence one knows that the length of [1,4]-soliton is linear with time and its end points will lie in a~line having slope $\pm \tan \chi$ (see Figs.~\ref{Fig4} and~\ref{Fig5}).
Furthermore, from~\eqref{x}, one gets that the [1,4]-soliton moves to the right if $\alpha_4<\frac{\pi}{3}$, and
moves to the left if $\alpha_4>\frac{\pi}{3}$.
In particular, if $\alpha_4=\frac{\pi}{3}$ or by~\eqref{co}
\begin{gather}
3+(\tan \varphi_C)^2 \big[3\kappa^2 (\tan \varphi_C)^2 -\kappa^2- 8 \kappa-1\big]=0.
\label{pt}
\end{gather}
then [1,4]-soliton's length is increasing along the~$y$-axis.
When $\kappa=1$ (or $\alpha_3=0$), one has $A=\frac{\epsilon}{2}$ by~\eqref{pt} and $\alpha_4=\frac{\pi}{3}$.
In this special case, the soliton is f\/ixed.
It is dif\/ferent from the KP-(II)
case~\cite{ko6, koo}.

\section[Relations with $V$-shape initial value waves]{Relations with $\boldsymbol{V}$-shape initial value waves}
\label{section3}
In this section, we investigate some relations with the $V$-shape initial value wave for the Novikov--Veselov
equation~\eqref{NV},~$\epsilon$ being f\/ixed, as compared with the KP-(II) case~\cite{ko1, ko5,ko6, koo}.
The main purpose is to study the interactions between line solitons, especially for the meaning of the critical
angle~$\varphi_C$.

Recalling the one-soliton solution~\eqref{on} and~\eqref{am}, one considers the initial data given in the shape of~$V$
with amplitude~$A$ and the oblique angle $\varphi_I<0$ (measured in the clockwise sense from the $y$-axis):
\begin{gather}
A\sech^2\big[\sqrt{2A} \cos \varphi_I (x-\vert y \vert \tan \varphi_I)\big].
\label{v2}
\end{gather}
For simplicity, one considers $A\leq 2 \epsilon$.
We notice here the $V$-shape initial wave is in the negative~$x$-region.
The main idea is that we can think the initial value wave as a~part of Mach-type soliton~\eqref{oo} or $O$-type
soliton~\cite{jh2}, that is, $c=0$ in~\eqref{oo}.
In order to identify those soliton solutions from the~$V$-shape~\eqref{v2}, we denote them as $[i^+, j^+]$-soliton for $y \gg 0$
and $[i_-, j_-]$-soliton for $y\ll 0$.
Solitons for $y \to \pm \infty $ have by~\eqref{on}
\begin{gather}
A=2 \epsilon \sin^2 \frac{\alpha_{j^+}-\alpha_{i^+}}{2}=2 \epsilon \sin^2 \frac{\alpha_{i_-}-\alpha_{j_-}}{2},
\label{A}
\\
\varphi_I=\frac{\alpha_{j^+}+\alpha_{i^+}}{2}=-\left(\frac{\alpha_{i_-}+\alpha_{j_-}}{2}\right).
\nonumber
\end{gather}
Assume that $ i^+<j^+$ and $ i_->j_-$.
Then symmetry gives
\begin{gather}
\alpha_{i^+}=- \alpha_{i_-},
\qquad
\alpha_{j^+}=- \alpha_{j_-}.
\label{sy}
\end{gather}
Using the parameter~\eqref{ka} \cite{ko,jh2, mi}
\begin{gather*}
\kappa=\frac{\vert \tan \varphi_I \vert}{\sqrt{\frac{A}{2\epsilon-A}}}=\frac{\vert \tan \varphi_I \vert}{\tan
\varphi_C},
\end{gather*}
one can yield, noticing
that $\varphi_C=\frac{\alpha_{j^+}-\alpha_{i^+}}{2}=\frac{\alpha_{i_-}-\alpha_{j_-}}{2}=\arctan \sqrt{\frac{A}{2\epsilon-A}}$ from~\eqref{A},
\begin{itemize}\itemsep=0pt
\item $\kappa \geq 1 \Rightarrow \vert \varphi_I \vert \geq \varphi_C
\Rightarrow -\frac{\pi}{2} \leq \alpha_{i^+}<\alpha_{j^+}<\alpha_{j^-}<\alpha_{i^-} \leq \frac{\pi}{2}$
($O$-type),
\item $0< \kappa<1 \Rightarrow \vert \varphi_I \vert<\varphi_C
\Rightarrow -\frac{\pi}{2} \leq \alpha_{i^+}<\alpha_{j^-}<\alpha_{j^+}<\alpha_{i^-} \leq \frac{\pi}{2}$
(Mach-type).
\end{itemize}
We remark here that if $\kappa=1$ (or $\alpha_3=0 $), then it is of~$O$-type by~\eqref{a} and~\eqref{y}.
One can see that if the angle $\varphi_I$ is small, then an intermediate wave called the Mach stem ([1,4]-soliton)
appears.
The Mach stem, the incident wave ([1,3]-soliton) and the ref\/lected wave ([3,4]-soliton) interact resonantly, and those
three waves form a~resonant triple.
It is similar to the KP-(II) case~\cite{ko1}.

Let's compute the maximal amplitude of the Mach stem ([1,4]-soliton) for f\/ixed amplitude~$A$ and~$\epsilon$.
By~\eqref{a}, a~simple calculation shows that
\begin{gather}
\frac{d A_{[1,4]}}{d \kappa}=A\frac{2(\kappa+1)[1-\kappa (\tan \varphi_C)^2]}{[1+\kappa^2 (\tan \varphi_C)^2]^2}.
\label{ma}
\end{gather}
Hence one can see that when $\kappa=1/(\tan \varphi_C)^2$, that is,
\begin{gather*}
(\tan \varphi_C) (\vert \tan \varphi_I \vert)=1,
\end{gather*}
the Mach stem has the maximal amplitude.
Consequently, if
\begin{gather}
\varphi_C+\varphi_I=\frac{\pi}{2},
\label{an}
\end{gather}
then one obtains by~\eqref{ma}, recalling that $ 0< \kappa<1$ (or $\tan \varphi_C>1$, i.e., $A>\epsilon$),
\begin{gather}
A_{[1,4]}^{\text{max}}=A\left(1+\frac{1}{(\tan \varphi_C)^2}\right)=2 \epsilon<2A.
\label{2a}
\end{gather}
Therefore one sees that from~\eqref{ma},~$A$ and $\epsilon $ being f\/ixed,
\begin{itemize}\itemsep=0pt
\item $ 0< \kappa<\frac{1}{(\tan \varphi_C)^2}$, the amplitude $A_{[1,4]}$ (stem wave) is increasing;
\item $\kappa=\frac{1}{(\tan \varphi_C)^2}$ (or~\eqref{an}), the amplitude $A_{[1,4]}$ has the maximal value $2
\epsilon $;
\item $\frac{1}{(\tan \varphi_C)^2}<\kappa<1$, the amplitude $A_{[1,4]}$ is decreasing.
\end{itemize}
It is noteworthy that the maximal amplitude is independent of~$A$.
Also, we know that the maximal amplitude of Mach stem for NV equation is less than twice of the incident wave's one;
however, for the KP equation (shallow water waves), the Mach stem's amplitude can be four times of the incident wave's
one~\cite{ko6}.
This is the dif\/ferent point from the case of the KP equation.

On the other hand, one can see that for $\kappa>1$ ($O$-type) we have $ 0 \leq \frac{\alpha_{j^+}-\alpha_{i^+}}{2}
\leq \frac{\pi}{4}$ by~\eqref{sy}, that is, $A\leq \epsilon$.
Thus, if we choose~$A$ such that
\begin{gather}
\epsilon<A\leq 2 \epsilon,
\label{2}
\end{gather}
we get $\frac{\pi}{2}<\alpha_{i^-} \leq \pi $; therefore, under the condition~\eqref{2}, the initial value
wave~\eqref{v2} would develop into a~singular $O$-type
soliton by~\eqref{oo} ($c=0$) when $\epsilon $ is f\/ixed.
On the other hand, when $\vert \varphi_I \vert \leq \frac{\pi}{2}$, $A$ and $\kappa$ are f\/ixed, one can choose
\begin{gather*}
\epsilon=\frac{A}{2}\left[1+\left(\frac{\kappa}{\tan \varphi_I}\right)^2\right] \geq \frac{A}{2}.
\end{gather*}
Then we can obtain regular soliton solutions.

Finally, from~\eqref{x} one remarks that the [1,4]-soliton (stem wave) moves to the right if $\alpha_4<\frac{\pi}{3}$,
and moves to the left if $\alpha_4>\frac{\pi}{3}$.
The former case is dif\/ferent from the KP equation (shallow water waves); i.e., the stem wave moves with the same side of
incident wave for the KP equation.
On the other hand, if we replace the condition~\eqref{A1} by $A=A_{[3,4]}=A_{[2,1]} \leq 2 \epsilon$, then by~\eqref{am}
the [3,4]-soliton (the incident wave) has smaller amplitude than the [1,3]-soliton's one (the ref\/lected wave).
But this is not physically interesting.

\section{Concluding remarks}
\label{section4}

One investigates the Mach-type (or (3142)-type) soliton of the Novikov--Veselov equation.
The Mach stem ([1,4]-soliton), the incident wave ([1,3]-soliton) and the ref\/lected wave ([3,4]-soliton) form a~resonant
triple.
From~\eqref{2a}, we see that the amplitude of Mach stem is less than two times of the one of the incident wave, which is
dif\/ferent from the KP equation~\cite{ko6}; moreover, the length of the Mach stem is computed and show it is linear with
time~\eqref{y}.
On the other hand, one uses the parameter~$\kappa$~\eqref{ka} to describe the critical behavior for the $O$-type and
Mach-type solitons and notices that it depends on the the f\/ixed parameter~$\epsilon$.
We see that the amplitude~$A$ of the incident wave is small than $2 \epsilon $; furthermore, if $\epsilon<A<2\epsilon$,
then the soliton will be singular.
Now, a~natural question is: what happens if $A> 2 \epsilon $ when~$\epsilon$ is f\/ixed in~\eqref{NV}? Another question
is the minimal completion~\cite{koo}.
It means the resulting chord diagram has the smallest total length of the chords.
This minimal completion can help us study the asymptotic solutions and estimate the maximum amplitude generated by the
interaction of those initial waves.
A~numerical investigation of these issues will be published elsewhere.

\subsection*{Acknowledgements}

The author thanks the referees for their valuable suggestions.
This work is supported in part by the National Science Council of Taiwan under Grant No.
NSC 102-2115-M-606-001.

\pdfbookmark[1]{References}{ref}
\LastPageEnding

\end{document}